# Three-dimensional array foci of generalized Fibonacci photon sieves

**Junyong Zhang,[1] Jie Ke,[1,2] Jianqiang Zhu[1], and Zunqi Lin[1]**

[1]*Shanghai Institute of Optics and Fine Mechanics, Chinese Academy of Sciences, Shanghai 201800, China*

[2]*University of Chinese Academy of Sciences, Beijing 100049, China*

*zhangjy829@siom.ac.cn*

**Abstract:** We present a new kind of photon sieves on the basis of the generalized Fibonacci sequences. The required numbers and locations of axial foci can be designed by generalized Fibonacci photon sieves (GFiPS). Furthermore, the three-dimensional array foci can be controllable and adjustable by the optical path difference scaling factor (OPDSF) when the amplitude modulation is replaced with the phase modulation. Multi-focal technologies can be applied to nano-imaging, THZ, laser communications, direct laser writing, optical tweezers or atom trapping, etc.

## 1. Introduction

Focusing of x-ray and extreme ultraviolet (EUV) has many applications in physical and life sciences, such as high-resolution microscopy, spectroscopy, and lithography. Although a traditional Fresnel zone plate (FZP) can be used for this kind of focusing, it has inherent limitations [1-5]. Some aperiodic zone plates, generated by fractal Cantor set, have been proposed to overcome some of these limitations [6-8]. Another interesting mathematical generator of aperiodic zone plates is the Fibonacci sequences. Photonics is a potential field of applications for novel devices designed and constructed by using a Fibonacci sequence as a consequence of its unique properties. The focusing and imaging properties of Fibonacci optical elements, e.g., quasicrystals [9, 10], gratings [11-13], lenses [14-16], zone plates [17], optical lattices [18,19], etc., are studied in detail. In mathematics, many mathematicians have extensively studied the Fibonacci sequence and its various generalizations [20-22] in the past decades.

In 2001, Kipp et al. proposed a new concept - photon sieve (PS) [23], which is a FZP with the transparent zones replaced by a great number of completely separated pinholes to overcome the disadvantages of traditional zone plate. Several kinds of theoretical models [24,25] had been studied mathematically and experimentally [26-28] to design different kinds of photon sieves, such as fractal [19,30], compound [31], Zernike apodized [32], phase zone [33], spiral [34], square [35], and reflection photon sieves [36]. In our previous work, we proposed a bifocal modified Fibonacci photon sieve (MFiPS), produced by using the Fibonacci sequence with two different initial seeds [37], but the ratio of the two focal lengths is a fixed value. This aperiodic sequence has not only been used for the design of photon sieves, but may allow for new applications in ophthalmology, nanometer lithography, and weapons vision.

In this paper, we introduce the aperiodic generalized Fibonacci sequences into photon sieves to generate three-dimensional array foci. Section 2 presents a general mathematical description of generalized Fibonacci sequences. In section 3, we propose a method to produce multiple axial foci with equal intensity under the condition of amplitude-only generalized Fibonacci photon sieve (GFiPS). Compared with the MFiPS, it can not only generate the variant ratios of the two focal lengths, but freely adjust the absolute locations of focal spots. When GFiPS with amplitude modulation completely replaced with phase modulation, it can be used to produce three-dimensional array foci illustrated in section 4. Besides, the axial foci of phase-only GFiPS have many similar focusing properties as that of amplitude-only GFiPS. At the end of the paper, the main conclusions of this work are outlined.

## 2. Description of GFiPS

For the generalized Fibonacci sequences, the initial seeds and the corresponding linear recursion relation are given by

$$\begin{cases} F_1 = a, F_2 = b, (a,b \in N) \\ F_n = pF_{n-1} + qF_{n-2}, (p,q \in R, n \geq 3, n \in N) \end{cases} \quad (1)$$

The absolute value of one of the corresponding characteristic roots of the recursion relation can be defined as the limit of the ratio of two consecutive generalized Fibonacci numbers

$$\gamma = \lim_{j \to \infty} F_j / F_{j-1} \quad (2)$$

When (p,q)=(1,1), we can get the standard Fibonacci sequence. The characteristic equation is $\gamma^2-\gamma-1=0$, and the corresponding characteristic roots are -0.6180 and 1.6180.

For traditional FZP, the optical path difference (OPD) between two adjacent zones is equal to $\lambda/2$, where $\lambda$ is the incident wavelength. If the OPD $\lambda/2$ is replaced with $K\lambda/2$, we can get the following OPD in the case of a monochromatic plane wave incidence

$$\sqrt{r_m^2 + f_0^2} - f_0 = mK\lambda, (m \in N_+, K > 0) \quad (3)$$

Where $f_0$ is the expected focal length, wavelength $\lambda$; $r_m$ is the radius of the $m^{th}$ zone; the positive real number K denotes the OPDSF.

Similar to the production process of the MFiPS, now taking a generalized Fibonacci sequence as an example, whose initial seeds are $(G_1,G_2)=(2,3)$ and its encoded seeds are $(G^1,G^2)=(01,010)$, the recursion relation is $G_n=-2G_{n-1}+0.3G_{n-2}$, the first four-order switching sequence $G^4$ is 010010010010 while 1 and 0 denote the transparent and opaque zone respectively. As indicated in Fig. 1 (b), the number of opaque zones is 8 and the number of transparent zones is 4. Fig. 1(a) shows the standard FiPS with the encoded seeds (01,010), the first four-order switching sequence is 01001010, which has 3 transparent and 5 opaque zones.

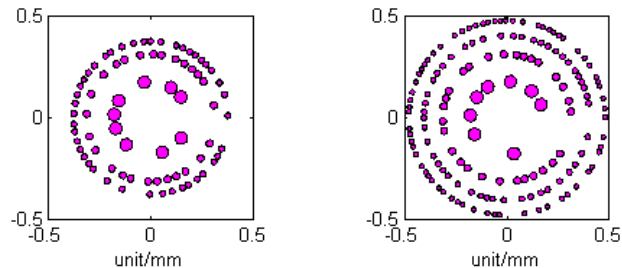

Fig.1. Schematic of GFiPS: (left) p=1, q=1, (right) p=-2, q=0.3

When a light source is incident on the pinhole, the diffracted field can be given by

Rayleigh-Sommerfeld diffraction integral

$$E_1(x,y,z) = \frac{-1}{2\pi} \iint E_0(\xi,\eta,0) t(\xi,\eta) \frac{\partial}{\partial z}\left[\frac{\exp(ikR)}{R}\right] d\xi d\eta \qquad (4)$$

Where $R=[(x-\zeta)^2+(y-\eta)^2+z^2]^{1/2}$, $E_0(\zeta,\eta,0)$ denotes the incident light, $t(\zeta,\eta)$ is the corresponding transmission function of each transparent pinhole. According to the linear superposition principle, the total diffracted field of GFiPS is the simple sum of those individual diffracted fields from different pinholes.

## 3. Axial multi-focal spots

In our previous work, we have investigated the bifocal GFiPS with different Fibonacci sequences in detail [38]. Based on this technology, here we provide two approaches to design the multi-focal GFiPS with axial equal intensity in this section. The simulation parameters are as follow: the wavelength $\lambda$ is 632.8nm, the expected focal length $f_0$ is 3.5cm, the encoded seeds ($F^1$, $F^2$) are (01, 010), and the recursion relations are $F_n=F_{n-1}+F_{n-2}$ and $G_n=-2G_{n-1}+0.3G_{n-2}$, respectively. Obviously, the latter characteristic equation is $\gamma^2+2\gamma-0.3=0$, and the characteristic roots are -2.1402 and 0.1402, respectively.

*3.1 Two axial foci*

In order to get two axial foci with equal intensity, the diameter of pinhole in the adjacent transparent zones is equal to the width of transparent zone to avoid partly overlapping, but the diameter of pinhole in the completely separate zones is equal to 1.16 times the width of zone. Figure 2 shows the axial intensity distribution against the distance. For the GFiPS with p=1 and q=1, the two focal lengths are 2.830cm and 4.579cm, respectively. As for the GFiPS with p=-2 and q=0.3, the two focal lengths are 2.567cm and 5.495cm, respectively. Their own ratios of the bifocal lengths are equal to 1.6180 and 2.1406, which approximates the modulus of its one characteristic root.

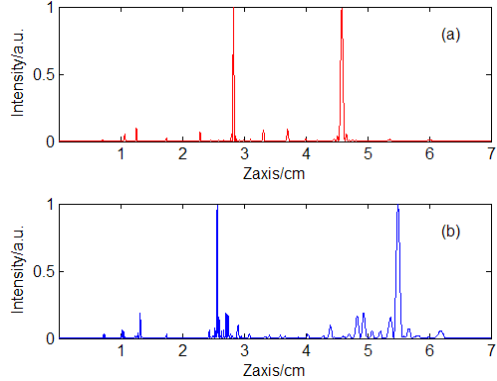

Fig.2. Axial intensity distribution against the distance by amplitude-only GFiPS with (a) p=1, q=1, (b) p=-2, q=0.3

*3.2 Multiple axial foci*

In this section, we discuss two approaches to get more than two foci on axis. The basis concept is to divide GFiPS into several pairs of independent regions which corresponds to the different switching sequences. Taking four axial foci into account under the condition of $f_2-f_1=f_4-f_3$, the recursion relation is $F_n=F_{n-1}+F_{n-2}$ in the first and the third regions, and in the second and the fourth regions, the recursion relation is $G_n=-2G_{n-1}+0.3G_{n-2}$. The total number of zones is 241 in each region.

Supposing that $f_2$ and $f_4$ correspond to the focal lengths of GFiPS with p=1 and q=1, naturally,

$f_1$ and $f_3$ correspond to another two focal lengths of GFiPS with p=-2 and q=0.3. Taking the GFiPS with p=1 and q=1 as the reference, we can reform the other GFiPS so as to satisfy the relation $f_2-f_1=f_4-f_3$. We have pointed out that the increase of focal length is proportional to the variation of OPD scaling factor. In this way, one approach is to reform the OPDSF; the other is to reform the expected focal length of the GFiPS with p=-2 and q=0.3. The corresponding solutions are (1) K=0.5 and $f_0$=2.1cm, (2) K=0.3 and $f_0$=3.5cm. The schematics of the two GFiPSs are shown in Fig.3. Figure 4 illustrates the axial intensity distribution against the distance by two different structures of amplitude-only GFiPSs. In the first and third regions, K=0.5 and $f_0$=3.5cm, and in the second and the fourth regions: (a) K=0.5 and $f_0$=2.1cm, (b) K=0.3 and $f_0$=3.5cm. The four focal lengths are 1.530cm, 2.830cm, 3.280cm, and 4.579cm, respectively. Obviously, $f_2-f_1=f_4-f_3\approx$1.3cm.

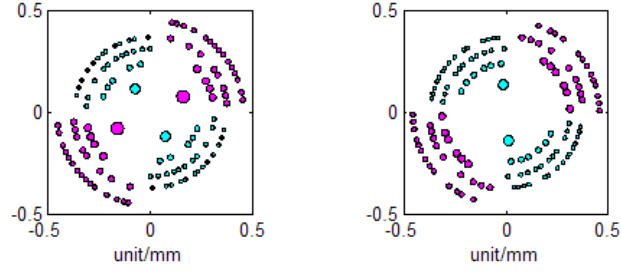

Fig.3. Schematic of amplitude-only GFiPS with four foci on axis, regions II and IV: (left) K=0.5 and $f_0$=2.1cm, (right) K=0.3 and $f_0$=3.5cm; regions I and III: K=0.5 and $f_0$=3.5cm.

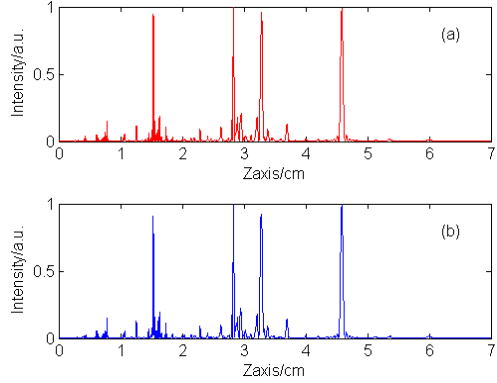

Fig.4. Axial intensity distribution against the distance by amplitude-only GFiPS, regions II and IV: (a) K=0.5 and $f_0$=2.1cm, (b) K=0.3 and $f_0$=3.5cm; regions I and III: K=0.5 and $f_0$=3.5cm

*3.3 The distinction between the two approaches mentioned above*

In section 3.2, we find that the two methods will have the same effect on the focal lengths in that there is a tiny perturbation for the value K of 0.5. Otherwise, if the OPDSF gravely deviates from the value of 0.5, there occurs an effect with a rather differently focusing structure on axis. The method by altering the expected focal length can't increase the number of the axial foci, while the number of axial foci will increase with the increase of the OPDSF, and the emerging foci satisfy the following relation

$$f_m = \frac{f_0 \times K}{\left[\frac{m-1}{2}\right] + \frac{\gamma \times \mathrm{mod}(m-1,2) + 1 \times \mathrm{mod}(m,2)}{\gamma + 1}}, f_m < f_{m-1} < \ldots < f_1 \quad (5)$$

Where [x] denotes the nearest integers less than or equal to x, mod(X,2) represents the modulus after division 2, and m denotes the $m^{th}$ axial focus.

Taking the GFiPS with p=-2 and q=0.3 for example, Fig. 5 and Table 1 shows the axial intensity distribution against the distance by amplitude-only GFiPS in the case of different OPDSF. From the simulation results, we know that the first two focal lengths is proportional to the value of K, but other focal lengths is not linearly varying with the value of OPDSF. No matter what, all the axial focal lengths almost obey the equation (5).

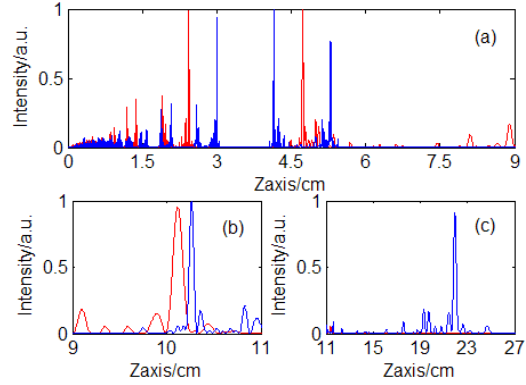

Fig.5. Axial intensity distribution against the distance by amplitude-only GFiPS with p=-2 and q=0.3 in the case of different OPDSF, (red line): K=0.92, (blue line): K=1.99

Table 1 Focal length of GFiPS with p=-2 and q=0.3 in the case of different OPDSF.

|  | K | $f_5$/cm | $f_4$/cm | $f_3$/cm | $f_2$/cm | $f_1$/cm | $f_1/f_2$ |
|---|---|---|---|---|---|---|---|
| simulation | 0.92 | - | - | 2.436 | 4.732 | 10.120 | 2.1386 |
|  | 1.99 | 2.999 | 4.148 | 5.293 | 10.260 | 21.940 | 2.1384 |
| based on Eq. (6) | 0.92 | - | - | 2.442 | 4.725 | 10.112 | 2.1401 |
|  | 1.99 | 3.004 | 4.142 | 5.283 | 10.220 | 21.872 | 2.1401 |

## 4. Three-dimensional array diffraction-limited foci

It's also worth pointing out that for an amplitude-only GFiPS, the transmission function of the transparent pinhole can be also considered as $\exp(in2\pi)$, where n is a positive integer. In section 3, we have obtained multiple foci on axis, but there is only one focal spot at the focal plane. Based on the concept of Fourier optics, the different phase stands for the different angular spectrum in the spatial domain. In this case, when the amplitude modulation is replaced by the phase modulation in the right way, the single focal spot may be split into multiple spots.

Similar to the production process of the amplitude-only GFiPS with multiple foci on axis, Figure 6 shows the schematic of the phase-only GFiPS with four equal regions. The value 0 still denotes the opaque pinhole, but the transparent pinhole is completely replaced with phase modulation. The phase is, in order, $\pi/2$, $\pi$, $3\pi/2$ and $2\pi$ in each region. The four intervals are, in order, $(-\pi/4,\pi/4]$, $(\pi/4,3\pi/4]$, $(3\pi/4,5\pi/4]$ and $(5\pi/4,7\pi/4]$.

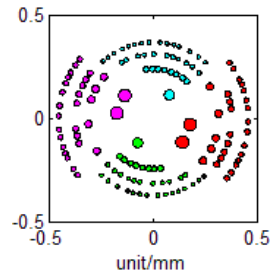

Fig.6. Schematic of phase-only GFiPS with four equal regions.

In Figure 7, the intensity distribution stands for the model corresponding to Figure 6. There are two focal spots at the each focal plane. The two focal spots along the y-axis are mainly produced by the first and the third region respectively, and at another two focal planes there are two focal spots along the x-axis which are mainly produced by the second and the fourth region. With this technology, we produce the three-dimensional array foci which is (1*2)*4. Meanwhile, the four focal planes satisfy the relation $f_2-f_1=f_4-f_3$. It's more important that the former two focal spots are orthogonal to the latter two focal spots.

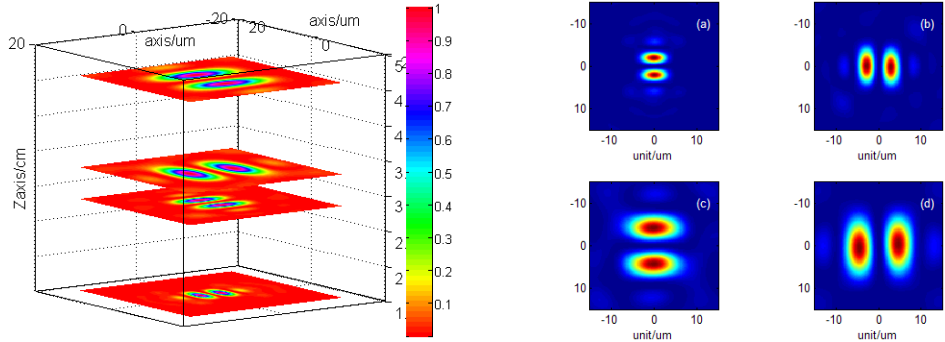

Fig.7. Intensity distribution (left): in 3D space, (right): at each focal plane.

To understand this technology well, we further give another example with (2*2)*3 array foci. For standard Fibonacci, the encoded seeds are (0,010), and there are 233 opaque zones and 89 transparent zones where the pinholes are located. The device can be divided into two parts along Cartesian coordinates, and the critical threshold is set to half width of device. Figure 8 shows the (2*2)*3 array focal spots. The transverse focal planes are located at 2.414cm, 3.164cm and 3.914cm on axis. Obviously, the interval between the adjacent focal planes is equal to 0.75cm.

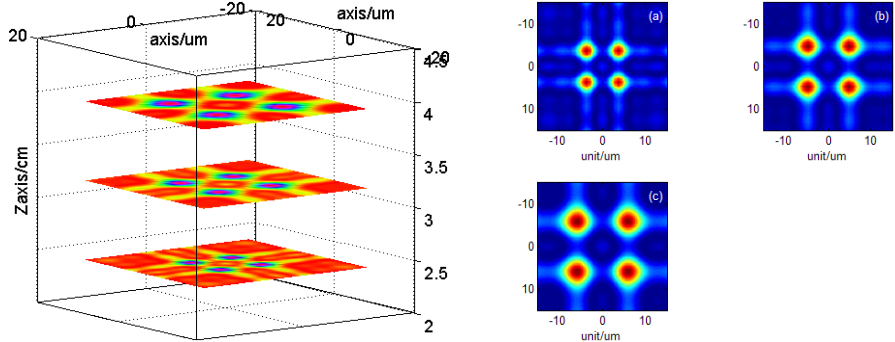

Fig.8. Array foci of phase-only GFiPS, (left): (2*2)*3 array, (right): intensity contours at each focal plane.

## 5. Conclusion
We extend the standard Fibonacci photon sieve to the generalized Fibonacci photon sieve. Multiple foci on axis with equal intensity have been obtained by merging the many different Fibonacci sequences in different areas. By replacing the amplitude modulation with the phase modulation for GFiPS, three-dimensional array foci have been successfully obtained. Multi-focal imaging technologies may offer a broad range of applications, such as laser communications, direct laser writing, optical tweezers or atom trapping and paralleled fluorescence microscope.

**Acknowledgements**
This research is supported by the National Natural Science Foundation of China (Grant Nos 61205212, 61205210).